# High-temperature stability and grain boundary complexion formation in a nanocrystalline Cu-Zr alloy

**Amirhossein Khalajhedayati[1], Timothy J. Rupert[1,2,3]**

1. Department of Chemical Engineering and Materials Science, University of California, Irvine, CA 92697, USA

2. Department of Mechanical and Aerospace Engineering, University of California, Irvine, CA 92697, USA

3. e-mail: trupert@uci.edu

**Abstract**

Nanocrystalline Cu-3 at.% Zr powders with ~20 nm average grain size were created with mechanical alloying and their thermal stability was studied from 550-950 °C. Annealing drove Zr segregation to the grain boundaries, which led to the formation of amorphous intergranular complexions at higher temperatures. Grain growth was retarded significantly, with 1 week of annealing at 950 °C, or 98% of the solidus temperature, only leading to coarsening of the average grain size to 54 nm. The enhanced thermal stability can be connected to both a reduction in grain boundary energy with doping as well as the precipitation of ZrC particles. High mechanical strength is retained even after these aggressive heat treatments, showing that complexion engineering may be a viable path toward the fabrication of bulk nanostructured materials with excellent properties.

## 1. Introduction

Refinement of average grain size to less than 100 nm results in a category of materials referred to as *nanocrystalline*. Unusual physical behavior and enhanced mechanical properties are often observed in nanocrystalline metals due to the high volume fraction of grain boundaries, motivating many research studies in recent years.[1,2] High corrosion, wear, and fatigue resistance are commonly reported, as are order of magnitude increases in strength.[3-9] For example, nanocrystalline Cu can achieve 10-15 times the strength of conventional microcrystalline Cu.[10] However, there are still major roadblocks that restrict the widespread implementation of nanocrystalline metals, mainly connected to the thermal stability needed to consolidate into bulk form.[11,12] The high grain boundary fraction gives a large driving force for coarsening and nanocrystalline materials exhibit limited thermal stability,[13] with grain growth even being reported at room temperature for pure nanocrystalline metals.[14]

The lack of thermal stability can be addressed in two ways: thermodynamic stabilization or kinetic stabilization of the microstructure. The former is achieved by reducing the excess grain boundary energy[15] while the latter is achieved by reducing the grain boundary mobility.[16] The effect of these two mechanisms can be summarized by Eq. 1:[4]

$$v = M \times P = M_0 \exp\left(\frac{-Q_m}{RT}\right) \times \frac{2\gamma}{r} \qquad (1)$$

where $v$ is the curvature driven velocity of a grain boundary during grain growth, $M$ is the grain boundary mobility, $P$ is the driving force for grain growth, $M_0$ is a mobility constant, $Q_m$ is an activation energy, $\gamma$ is grain boundary free energy per unit area, and $r$ is the radius of curvature. To obtain a stable microstructure, either $M$ or $P$ should be reduced, ideally to zero if stability is expected over long time periods. Kinetic approaches often rely on precipitation of second phase particles or solute drag that reduces the grain boundary mobility by applying a pinning pressure on grain boundaries.[17] However, based on Eq. 1, the mobility term follows an Arrhenius function and extremely high temperatures can ultimately make kinetic approaches impractical. In thermodynamic approaches, $\gamma$ should be reduced since the driving force for grain boundary migration is proportional to grain boundary energy. This is often done by doping grain boundaries with a solute that has a large atomic size mismatch with the solvent atoms. Grain boundary free energy is much less temperature dependent, and thermodynamic methods are philosophically more promising since they could stabilize a microstructure indefinitely.[18-20] The common theme between both approaches is that alloy systems are required for high stability, meaning that the doping of nanocrystalline materials should be studied in more detail

A number of theoretical models have been developed to predict candidate elemental combinations for thermally-stable nanocrystalline materials, most using readily available thermodynamic parameters.[17,21] For example, Murdoch and Schuh[22] developed a comprehensive model using a Miedema-type[23] approach for the estimation of grain boundary segregation enthalpy, with their model being independent of temperature and grain boundary solute concentration. These authors used this model to find the most thermally-stable binary alloys and confirmed previous experimental observations, such as excellent thermal stability in nanocrystalline W-Ti with a grain size of ~20 nm[15] and a combination of phase separation and grain growth in nanocrystalline Cu-Ag.[24] In a similar manner, to predict the stability of many nanocrystalline systems due to grain boundary segregation, Darling et al.[25] created stability maps showing the minimum solute composition required to reduce excess grain boundary energy to zero at different grain sizes and temperatures. Both of these studies serve to guide researchers in their selection for system candidates when engineering thermally-stable nanocrystalline alloys through grain boundary segregation.

Besides having an important role in reducing grain boundary energy, grain boundary segregation can drive the formation of new interfacial structures at the grain boundaries.[26] Initially, Hart[27] suggested that grain boundary segregation can lead to the nucleation of grain boundary phases that could explain embrittlement in steel. Subsequently, Clarke and Thomas[28] were able to directly show evidence of interfacial phases in a MgO-doped $Si_3N_4$ ceramic using high resolution transmission electron microscopy (TEM). Many studies have since reported on the effects of these grain boundary structures on the overall physical and mechanical properties in metals and ceramics, such as the being cause for solid state activated sintering and abnormal grain growth.[29-37]

Recently, Tang et al.[38] introduced the term *complexion* to differentiate grain boundary "phases" from traditional bulk phases. The main reason for this convention is that complexions depend on their abutting grains and cannot exist as separate entities.[26] However, there are some similarities which are useful for materials design. Grain boundary complexions, like bulk phases, can be described with thermodynamic parameters and can transition between complexion types due to changes in variables such as temperature, pressure, local chemistry, or grain boundary character.[26] Dillon et al.[39] categorized complexions into six types based on the grain boundary thickness and solute segregation: single layer segregation (Type I), clean or undoped grain boundaries (Type II), bilayer segregation (Type III), multilayer segregation (Type IV), nanoscale intergranular films (Type V), and wetting films (Type VI). Complexion types I-V are only in equilibrium when sandwiched between abutting crystals and imply a reduction in grain boundary energy, meaning they may be present during the thermodynamic stabilization described by Murdoch and Schuh[22] and Darling et al.[25] Alternatively, complexion type VI can be in equilibrium at an interface or by itself as a standalone phase. The different complexions can also be grouped in terms of their local structure. For example, complexion types I-IV retain crystalline order in the interface and can be described as ordered complexions. In contrast, complexion types V and VI are often disordered, in which case they can both be broadly classified as amorphous intergranular films.

    Complexions were first observed in ceramic systems, but have since also been found in metallic materials. For example, it was recently discovered that the embrittlement of Cu[40] and Ni[41] in the presence of Bi is due to formation of a bilayer complexion with stretched atomic bonds that are easily broken. Segregation of Ga in Al grain boundaries is another example of embrittlement due grain boundary complexion formation.[42] However, complexions are not always deleterious to mechanical properties. Molecular dynamic simulations by Pan and Rupert[43,44] on Cu bicrystals doped with Zr showed that complexions in the form of amorphous intergranular films (complexion types V or VI) can increase the damage tolerance of a grain boundary by absorbing multiple dislocations and acting as an efficient sink for dislocations. These authors also showed that thicker amorphous intergranular films are capable of absorbing more dislocations and delaying intergranular void formation to higher shear strains.[43] Because nanocrystalline metals with grain size above 10 nm plastically deform through grain boundary dislocation emission and absorption,[45] grain boundary complexions should have a major impact on the ductility and toughness of these materials. In fact, this hypothesis is supported by the work of Wang et al.[46] on nanocrystalline Cu/Cu-Zr nanolaminates. Although they were not complexions created by thermodynamic driving

forces, the thin amorphous layers that were added between nanocrystalline Cu layers were found to increase ductility when compared to a monolithic nanocrystalline Cu film.

It is apparent that grain boundary complexions have the potential to dramatically alter the behavior of nanocrystalline materials, and could potentially be formed when high levels of grain boundary doping are present and elevated temperatures are experienced. Consequently, complexions may alter the thermal stability of these materials as well. To date, little attention has been given to grain boundary complexion formation and structural transitions in doped nanocrystalline metals. In this study, we report on the annealing of a nanocrystalline Cu-Zr alloy created by high-energy ball milling, with a focus on studying the segregation of dopants to grain boundaries and the formation of complexions as a function of temperature. Our results indicate thermal stability in nanocrystalline samples annealed at temperatures as high as 98% of the solidus temperature for a period of 1 week. Upon closer inspection, grain boundary complexions were observed, with the level of interfacial disorder and grain boundary thickness increasing as temperature increased.

## 2. Materials and Methods

Nanocrystalline Cu-3 at.% Zr alloy powders were produced using a SPEX 8000M high-energy ball mill. Cu was chosen as the base element since nanocrystalline Cu systems have been studied extensively in the literature.[4] Zr was then picked as the doping element for several reasons. Cu and Zr have a large atomic size mismatch and different crystal structures, so Zr has a high tendency to segregate to the grain boundaries in this mixture. Such behavior is also predicted by the fact that Zr has very limited solid solubility in Cu, only 0.12 at.% at the eutectic temperature.

Furthermore, Cu and Zr have been shown to have good glass forming ability, giving the possibility of amorphous intergranular film formation.[47,48]

A 99.99% pure and -170+400 mesh Cu powder from Alfa Aesar was mixed with a 99.7% pure and -50 mesh Zr powder from Micron Metals in a hardened steel vial. The ball milling machine was placed inside of a glove box in a 99.999% pure Ar atmosphere with a purifier system that extracts residual oxygen, to avoid oxidation during milling. Examination with energy-dispersive X-ray spectroscopy (EDS) after milling shows no measurable oxygen content, meaning any oxidation is below the resolution of the technique, or ~0.1 at.%. Stearic acid in the amount of 1 wt.% was added to the mixture as a process control agent to prevent cold welding. The powder mixture was milled for 10 h with a ball-to-powder ratio of 10:1 using hardened steel milling media. To prepare the samples for heat treatment, each powdered sample was sealed under vacuum in a high purity quartz tube. Annealing treatments were performed with a vertical tube furnace and the samples were quickly quenched by dropping into a water bath in less than 1 s, in order to preserve any high temperature interfacial structures. Samples were annealed at 550 °C, 750 °C, 850 °C, and 950 °C. Three samples were made at each temperature by annealing for 1 h, 24 h (1 day) and 168 h (1 week). All the samples were analyzed with X-ray diffraction (XRD) before and after annealing using a Rigaku Ultima III diffractometer with a D/teX Ultra 250 1D detector for grain size measurement and phase identification. XRD grain size was measured using the (111) peak of Cu and the Scherrer equation.[49] It should be noted that this method ignores any microstrain that might be in the powders, but this means that any measured grain growth will be an upper limit or worst case scenario as some of the reduction in peak width will be associated with relaxing microstrain. As will be described shortly, we also perform TEM analysis to confirm general trends in grain size. Consequently, an error bar of ±15% has been added to the measured XRD grain size of each sample to account for uncertainties that results from theoretical assumptions.[50] Each powder sample was then mounted, without compaction, using a cold conductive epoxy system on a standard scanning electron microscopy stub and polished to a mirror surface finish.

TEM samples were made from individual powder particles with a focused ion beam using a FEI dual beam Quanta3D microscope. Each TEM sample was polished with a 5 kV $Ga^+$ ion beam to reduce the beam damage. Bright field and dark field TEM images, as well as selected area electron diffraction (SAED) patterns, were taken using a Philips CM-20 operating at 200 kV. Average grain sizes from TEM were found by measuring at least 100 grains and taking the equivalent circular diameter. Energy-dispersive X-ray spectroscopy (EDS) and scanning TEM imaging were performed using an FEI Titan operating at 80 kV, while high resolution TEM imaging was carried out with the same instrument at 300 kV. Fresnel fringe imaging was employed to find grain boundary films and identify edge-on grain boundaries.[51] Finally, hardness measurements were obtained with an Agilent nanoindenter equipped with a Berkovich tip. A constant loading rate of 1 mN/s and an indentation depth of 400 nm were employed. At least 20 measurements were taken for each sample and annealing condition.

3. Microstructure characterization

3.1. *Grain size and grain boundary segregation*

The XRD grain sizes of the annealed samples are shown in Fig. 1 as a function of annealing time. The grain size of the as-milled sample (18 nm) is also plotted as a straight orange line for reference in the same figure, with error bars denoted by orange shading. Small increases in the average grain size are found after annealing the samples for 1 h at 550 °C, 750 °C, and 850 °C, followed by slow but constant growth at longer times. Even after annealing these samples for a period of 1 week, the grain sizes remain below 35 nm. Elemental Cu has a relatively low melting temperature of 1084 °C, but the solidus temperature where Cu-3 at.% Zr begins to melt is even lower at 972 °C. Therefore, annealing temperatures of 550 °C, 750 °C, and 850 °C are high for this alloy, being 66%, 82%, and 90% of the solidus temperature, respectively. Increasing the annealing temperature to 950 °C, or 98% of the solidus temperature, leads to higher levels of grain growth compared to the other annealing conditions. However, the material remains nanocrystalline even after annealing for a period of 1 week, with an average size of ~54 nm.

To confirm the trend observed with XRD, TEM analysis was also performed. Fig. 2 shows bright field TEM images of the samples annealed for 1 h. All of the samples have equiaxed grain structures and no abnormal grain growth was detected. The insets in the figure show SAED patterns for each sample. Cu FCC rings are clearly visible with the addition of faint rings of a second phase, most easily seen inside of the (111) Cu diffraction ring. Further analysis showed that ZrC particles are formed during annealing, which will be discussed in more detail in Section 3.3. To compare the XRD and TEM grain sizes, cumulative distribution functions of grain size are plotted in Fig. 3. XRD gives a slightly smaller grain size measurement in each case, likely as a result of neglecting microstrain in the Scherrer equation.[49] However, a similar trend in grain growth is observed by both characterization methods and the samples remain nanocrystalline.

To understand the cause of the thermal stability of this alloy, EDS line profile analysis using scanning TEM was conducted to probe the Zr distribution within the microstructure, as shown in Figs. 4 and 5. Care was taken to make sure the concentration profiles were taken only from the FCC phase and not influenced by any ZrC particles in the microstructure. Interaction of the electron beam with the ZrC particles results in a sharp increase in the Zr concentration to greater than 40 at.%, which was not observed here. The EDS data presented here should be considered semi-quantitative, not an exact measure of concentration at a point. As opposed to microcrystalline samples, EDS data obtained from nanocrystalline TEM samples may contain information from more than one grain through the thickness of the sample. The beam interaction volume may also lead to some slight spatial averaging on a nanometer scale.

Fig. 4a shows a scanning TEM image and line profile scan of the sample annealed at 550 °C for 1 h. The concentration of Zr in this sample reaches maximas of about ~5-6 at.%, and a section of the profile is magnified directly below. The fluctuation of the Zr content in the sample roughly matches the observed grain size, and the locations of two obvious grain boundaries are labeled. The concentration of Zr reaches zero at several points, a sign that some local regions are depleted of Zr, but grain interiors are still doped. A similar behavior is observed when increasing the annealing temperature to 750 °C, with one noticeable difference being the higher amplitude of Zr peaks to maximas of ~7-8 at.% (Fig. 4b). In addition, some large areas are fully depleted of Zr, providing evidence of full segregation at the grain boundaries. Previous works on characterization of solute distribution in nanocrystalline W-Ti[15] and Ni-W[52] only showed partial segregation of solutes in those systems, with crystal interiors that were still heavily doped. Here, full segregation of Zr atoms to the grain boundaries of nanocrystalline Cu is observed for annealing temperatures at and above 750 °C.

Fig. 5 shows scanning TEM images and EDS line profile scans of the samples annealed for 1 h at 850 °C and 950 °C. The concentration of segregated Zr at the grain boundaries slightly increased to ~10 at.% in most cases and the spacing between the Zr peaks is also wider, showing the effect of grain growth. The spacing between the maximums in the profile is closely related to the grain size of the samples. A high concentration of Zr was observed at some grain boundaries, exceeding 20 at.% Zr in rare cases such as the example denoted by an arrow in Fig. 5a. These observations were not common but did occur occasionally, showing that some grain boundaries may have been more heavily doped than average. Again, grain interiors appear to be depleted of Zr.

*3.2. Grain boundary structure*

As discussed in the introduction, a combination of high segregation levels and high temperature can lead to changes in grain boundary structure, namely transitions between different complexion types. Fig. 6 shows high resolution TEM images of representative interfacial structures for the samples annealed at 550 °C and 750 °C for 1 h. In both cases, the grain boundaries are atomically sharp, with no obvious phase-like structure at the interface, and show a low-energy configuration which is an ordered and continuous grain boundary structure that may contain a periodic dislocation array. The structure of the triple junction in Fig. 6b is another example of ordered interfacial regions in these samples.

Increasing the annealing temperature to 850 °C and 950 °C, distinct grain boundary structures begin to emerge. Fig. 7a shows a 2.6 nm thick amorphous intergranular film between two grains in the sample annealed at 850 °C. Fig. 7b shows the existence of another amorphous intergranular film with sub-nanometer thickness (0.8 nm) within the same sample. The thickness of these films is constant across a given grain boundary, suggesting that they are in equilibrium with the abutting grains.

Fig. 7c shows a high resolution TEM image of a grain boundary in the sample annealed at 950 °C. A relatively wide amorphous intergranular film with a thickness of 4.1 nm can be seen. Directly below this figure are the Fast Fourier Transform (FFT) images of the grain boundary and the two abutting grains. The FFT of two grains show periodic spots around the center of the image, as expected from crystalline grains. On the other hand, the FFT of the grain boundary film shows no features, representative of a disordered amorphous layer. Fig. 7d shows another high resolution TEM of an amorphous intergranular film for the 950 °C sample with a thickness of 2.9 nm. Again, the interface of this film with the abutting grains is sharp and the thickness of this film does not change along the length of the grain boundary. Overall, amorphous intergranular films in the 950 °C sample with thicknesses ranging from 0.5 nm to 5.7 nm were observed. It is also noteworthy that amorphous intergranular films were observed more frequently in the sample annealed at 950 °C than the 850 °C sample. However, it is difficult to be quantitative about the fraction in each case because not all grain boundaries in a given area of a TEM sample can be adequately investigated with high resolution TEM, due to limitations on finding edge-on grain boundaries.

### 3.3. Second phase precipitation

Precipitation of ZrC particles was detected after annealing the samples, as was shown in the insets of Fig. 2. Initially, a small volume fraction of $ZrH_2$ particles is detected from the XRD pattern in the as-milled sample (Fig. 8a). It is likely that the $ZrH_2$ precipitates were formed during milling due to reaction of excess Zr with H atoms, likely coming from the stearic acid used as the process control agent. Although stearic acid has a great deal of both H and C, $ZrH_2$ has a lower enthalpy of formation compared to ZrC and the mobility of H is higher than C in Zr. Annealing at higher temperatures led to formation of ZrC particles, as shown in the XRD pattern (Fig. 8a) and the SAED pattern (Fig. 8b) of the sample annealed for 950 °C. This likely occurs because annealing breaks carbon-carbon bonds in stearic acid and freed C atoms, which then react with $ZrH_2$ at high temperatures to form ZrC.[53] The absence of $ZrH_2$ and the presence of ZrC was observed for all annealed samples.

Fig. 9a-c shows dark field images of the samples annealed for 1 h at three different temperatures. ZrC particles are homogenously dispersed throughout the microstructure and were found both in the grain interiors as well as on the grain boundaries. Comparison of Fig. 9a and c demonstrates that ZrC particles grew larger in size after annealing at higher temperatures. Fig. 9d and e show a zoomed dark field TEM image and a high resolution TEM image, respectively, of a ZrC particle in the 750 °C sample. The high resolution TEM image shows that this ZrC particle is sitting along a grain boundary. To further analyze the ZrC particles, XRD particle size of ZrC is plotted as a function of annealing time in a semi-log plot in Fig. 10. The trend in ZrC particle growth is similar to that observed for grain size. Annealing at 850 °C and below for a period of 1 week results in small particles with average particle sizes of ~10 nm or less. By increasing the annealing temperature to 950 °C, ZrC particle size increases, to a maximum of ~25 nm after 1 week. The observed particle growth can be due to diffusion of Zr atoms along the grain boundaries to reach ZrC particles or by coalescence of smaller particles as the grain boundaries migrate.

## 4. Discussion

### 4.1. *Grain size stability*

The results of the XRD and TEM analysis show significant grain size stabilization in this alloy, even after annealing under extreme conditions. This level of thermal stability is especially notable, because it surpasses the temperatures and times necessary to obtain fully dense nanocrystalline Cu through sintering processes. Sintering of nanocrystalline Cu is often performed in the temperature range of 400-800 °C for a period of 1-24 h, depending on the type of sintering technique.[54,55] Additionally, our results agree with the recent study by Atwater et al.,[56] who showed that a nanocrystalline grain size can be retained after annealing at temperatures as high as 800 °C for cryogenically ball milled Cu-Zr without the addition of process agents. These authors attributed their results to the segregation of Zr atoms to the grain boundaries at low temperatures, while also showing that second phase intermetallic particles begin to from at temperatures higher above 800 °C to help stabilize the grain boundaries through Zener pinning. Since both grain boundary segregation and second phase precipitates are present in our study as well, stability combination of thermodynamic and kinetic stabilization may be responsible for the observed retardation of grain growth.

Based on the study by Darling et al.,[25] a true thermodynamic equilibrium for nanocrystalline Cu-Zr with an average grain size of 25 nm at 650 °C is possible but requires a global concentration of 6 at.% Zr, higher than that used here. However, according to the molecular dynamic simulations and thermodynamic studies of Millet et al.,[57] if the ratio of atomic radius of a solute to atomic radius of Cu is equal to 1.25, a global dopant concentration of 0.5 at.% is enough to reduce the excess grain boundary energy to zero in a nanocrystalline Cu alloy with a grain size of 20 nm. By considering the atomic radii of Zr (159.7 pm) and Cu (127.6 pm),[58] the ratio of atomic radius of Zr to Cu is 1.25, meaning that a dilute concentration of Zr should be able to stabilize the grain size of nanocrystalline Cu-Zr. Our results showed a significant degree of solute segregation to grain boundaries, so we expect that thermodynamic concerns play a major role in stabilizing our nanocrystalline Cu-3 at.% Zr alloy, in agreement with the prediction of these studies. However, such models tend to ignore kinetic effects, such as those associated with desegregation of solutes from the grain boundary and formation of second phase particles.

While we do not observe desegregation, second phase particles do exist and could have an effect on the thermal stability of Cu-3 at.% Zr as well. The effect of ZrC and $ZrO_2$ particle pinning on the thermal stability of a ball milled nanocrystalline Cu was studied by Morris and Morris.[59] After adding 10.8 vol.% of second phase particles in that study, annealing treatments of 900 °C for 1 h resulted in a grain size of 73 nm. In another study, the effect of Ta particles on thermal stability of nanocrystalline Cu-10 at.% Ta was confirmed and an average grain size of 111 nm was reported after annealing the microstructure for 4 h at 900 °C.[60] Therefore, it is clear that addition of second phase particles can hinder grain growth through Zener grain boundary pinning mechanisms. Our peak intensity ratio date from XRD shows that our nanocrystalline Cu-3 at.% Zr alloy could contain 2-4 vol.% of ZrC particles, depending on the annealing temperature and time. An average grain size due to grain boundary pinning by second phase particles can be estimated by a simple Zener equation:[61]

$$D_{Zener} = K_{Zener} \frac{d_p}{f^m} \qquad (2)$$

where $d_p$ and $f$ are secondary phase particle size and volume fraction respectively, and $K_{Zener}$ and $m$ are constants which can be taken as 0.17 and 1 if the volume fraction of second phase particles is less than 0.05.[61] By taking $f = 0.02$ and $d_p = 5$ nm for the sample annealed at 550 °C for 1 h, Eq.

2 gives $D_{Zener}$ = 42.5 nm, which is approximately twice as large as the measured grain size of this sample. This calculation suggests that particle pinning cannot fully account for the observed thermal stability in nanocrystalline Cu- 3 at.% Zr. Consequently, both thermodynamic stabilization through grain boundary segregation and kinetic stabilization through Zener pinning seem to be operating here.

*4.2. Complexion formation in nanocrystalline Cu-Zr*

Our TEM results showed distinct grain boundary structures in nanocrystalline Cu-3 at.% Zr after annealing at different temperatures. As previously shown by Dillon et al,[39] each grain boundary state has a unique property set that can cause a dramatic change in the overall material properties, motivating further discussion. At the lowest annealing temperature of 550 °C, only ordered grain boundary structures are observed and EDS results show that all of the grain boundaries are doped with Zr. Therefore, the observed grain boundary structures at this temperature are ordered complexions of type I, III, or IV. In general, differentiation of single layer, bilayer, and multilayer segregation is extremely difficult and requires an aberration-corrected scanning TEM.[62,63] Additional difficulty arises because accurate imaging of these thin complexions requires a perfect edge-on imaging condition and even a small amount of misalignment can lead to overlapping of lattice fringes from the neighboring grains, effectively covering the thin complexions from view.[64] Ordered complexions are more energetically favorable at lower temperatures and lower dopant compositions, having a high degree of crystalline order at the interface compared to other complexion types.[65] Furthermore, ordered complexions have been shown to lower grain the boundary mobility in MgO-doped $Al_2O_3$ and cause embrittlement in Bi doped Cu.[39,66] It is important to note that these grain boundary structures represent the thermodynamically stable configurations that form at that annealing temperature, since fast quenching was employed. However, the conventional equilibrium phase diagram (Fig. 11)[67] for binary Cu-Zr alloy has no information about complexions and only a mixture of Cu and $Cu_9Zr_2$ should be thermodynamically-stable at this temperature, although such a mixture is not seen here.

Further increasing the annealing temperature to 750 °C caused an increase in the degree of grain boundary segregation compared to 550 °C, but again ordered complexions were the only grain boundary structures that were identified. A study by Dillon et al.[39] on complexion formation in $Al_2O_3$ showed that type V and VI complexions can start to form at temperatures at and above ~0.71 of melting point of $Al_2O_3$, which would translate to 615 °C for Cu-3 at. % Zr. The lack of amorphous intergranular films at 750 °C may suggest that the formation of quasi-liquid films requires higher homologous temperatures in metallic alloys.

Nanometer thick quasi-liquid intergranular films were observed at 850 °C. These grain boundaries resemble the intergranular grain boundary films which are formed in coarse-grained polycrystalline ceramics and metals,[68] but have not been found in nanocrystalline metals previously. It should be noted that the annealing temperature here is below the eutectic temperature (Fig. 11b), and therefore this complexion cannot be a meta-stable bulk liquid phase. Additionally, the thickness of amorphous grain boundaries films was constant along a given interface, confirming the equilibrium state of the amorphous film with the neighboring grains. All of the mentioned conditions match the criteria of a Type V complexion. The maximum thickness of the observed complexion Type V at 850 °C has been marked in Fig. 11a. Recently, Luo and coworkers developed a framework to predict the equilibrium grain boundary film thicknesses at sub-solidus temperatures in binary alloys:[69]

$$\lambda \equiv -\Delta\gamma / \Delta G_{amorph} \quad (3)$$

where $\lambda$ is a thermodynamic parameter proportional but not equal to film thickness, $\Delta\gamma$ is the difference between the excess energy of a clean grain boundary and the crystalline-amorphous interfaces, and $\Delta G_{amorph}$ is the volumetric free energy of formation of undercooled liquid. These authors showed that $\lambda$ is related to temperature, mainly because the free-energy penalty for forming an undercooled liquid decreases with increasing temperature, so increasing annealing temperature should result in a thickening of any amorphous grain boundary films. As a result, annealing at 950 °C leads to the formation of thicker amorphous complexions. The maximum thickness of any amorphous complexion observed at this temperature, 5.7 nm, is marked in the Fig. 11a. It is important to note that we also observed ordered grain boundaries in both the 850 °C and 950 °C samples, so not all boundaries are covered with amorphous intergranular films and a distribution of complexions types is observed. This distribution is likely related to the fact that grain boundary free energy depends not only on temperature and chemical potential, but also on the starting grain boundary structure. Interfaces have five degrees of freedom, related to grain misorientation and grain boundary normal, which can also influence the energy and thermodynamics of the interfaces.

While the amorphous films in the sample annealed at 850 °C can only be Type V complexions, it is possible that the amorphous intergranular films observed in the 950 °C sample are a combination of Type V and Type VI. Although the annealing temperature is below the solidus temperature for the global composition used here, it is above the eutectic temperature, or the lowest temperature for the binary alloy where any liquid phase is in equilibrium (Fig. 11b), and it is important to recall that some grain boundaries showed levels of Zr segregation to >20 at.%. If the local composition at the grain boundary exceeds 37.5 at.% Zr, it is possible that a wetting film can nucleate at 950 °C. Although it is difficult to definitively classify the amorphous films in this sample, the two competing complexion types has different characteristics. Complexion Type V is in equilibrium, so a uniform thickness should be maintained along the entirety of a given grain boundary. On the other hand, complexion Type VI is a bulk phase, so it can change thickness along the length of an interface based on the available concentration of dopant atoms. All of the amorphous films observed in the 950 °C sample showed no measureable change in thickness here, and the wetting films that have been observed in other materials are often much thicker, often being at least 10 nm thick. It is also worth highlighting that grain boundaries with Zr content above ~10 at.% were rarely observed, meaning it is likely that most if not all of the observed amorphous films were Type V.

Lastly, our grain growth results can be revisited to understand the effect of grain boundary complexion structure on grain size stabilization. To obtain a clear picture, only the highest and lowest annealing temperatures are compared. At 550 °C, grain growth is slow and the error bars on average grain size still overlap between annealed and as-milled after 1 week. Only ordered complexions were found in this sample, and previous studies in $Al_2O_3$ showed that this type of complexion has the lowest mobility of all types.[39] In contrast, type V and VI complexions have been shown to dramatically speed up diffusion and cause abnormal grain growth or activated sintering.[39,70] Observation of both a higher grain growth and formation of type V or VI complexions in the sample annealed at 950 °C shows a correlation between the thermal stability and complexion type. Since grain boundaries can act as short-circuit paths for diffusion, it is also worth noting that the interfaces are thicker when the higher level complexions are formed at 950 °C.

*4.3.   Effect of structural features on hardness*

Thermal stability is essential in nanocrystalline metals because it retains the properties that motivated research in nanocrystalline metals in the first place. However, the impact of segregation, complexions, and precipitates on mechanical properties should be considered as well. Nanoindentation hardness data for each sample is plotted as a function of annealing time in Fig. 12a. For comparison, the hardness value of a pure Cu sample with an average grain size of 22 nm, made with the same processing conditions, is also included in the same plot as a blue line with error bars denoted by light blue shading. From the figure, it is clear that the Cu-Zr alloys exhibit hardness values that are always higher than the as-milled pure Cu, even after annealing at extremely high temperatures for very long times. A maximum hardness, ~4.4 GPa, is reached in the samples annealed at 550 °C for 24 h and 750 °C for 1 h. Since annealed samples have larger grain sizes compared to the pure nanocrystalline Cu sample, the strength of the alloy should have decreased if grain size was the only consideration, following the Hall-Petch relationship:

$$\sigma_{Hall-Petch} = \sigma_o + \frac{k}{\sqrt{d}} \qquad (4)$$

where $\sigma_o$ is a frictional stress and $k$ is a constant, taken as 25.5 MPa and 0.11 MN/m$^{3/2}$, respectively, for Cu.[71] However, the opposite is observed, with the larger grain size annealed samples show higher strength. Therefore, other hardening mechanisms must be at work in this alloy, with grain boundary segregation and precipitation hardening being prime candidates.[72]

Some studies suggest that Orowan hardening, or the bowing of dislocations between second phase particles, is mainly responsible for the observed increase in hardness of the nanocrystalline metals that contain precipitates.[71] The increased in yield stress due to an Orowan hardening mechanism can be estimated by:[73]

$$\Delta\sigma_{Orowan} = \frac{0.13Gb}{\lambda} \ln\left(\frac{d_p}{2b}\right) \quad (5)$$

where $G$ and $b$ are the shear modulus and Burgers vector, respectively, of Cu, $d_p$ is the particle size, and $\lambda$ is the interparticle spacing which is estimated by:[73]

$$\lambda \approx d_p \left[(\frac{1}{2f})^{1/3} - 1\right] \quad (6)$$

where $f$ is the volume fraction of precipitates. The combined effect of Orowan hardening and Hall-Petch hardening can then be calculated by assuming $H \sim 3\ \sigma_y$:[74]

$$H_{total} = 3\ (\sigma_{Hall-Petch} + \Delta\sigma_{Orowan}) \quad (7)$$

Taking $G$ = 48 GPa, $b$ = 0.255 nm, and assuming a constant volume fraction of $f$ = 0.04 for all the annealing conditions, which is the highest value found in any sample and serves to provide an upper bound on Orowan strengthening contributions, the Hall-Petch and Orowan hardness values were calculated and are plotted in a stacked bar chart in Fig. 12b. The combined Hall-Petch and Orowan hardening effects almost fit some of the lower temperature annealing conditions. However, as the annealing temperature increases, the discrepancy between the simple grain size-precipitate model and our experimental results widens, suggesting that additional mechanisms begin to significantly contribute to the hardness. It is also important to remember that we have used an upper bound for the volume fraction of secondary phase particles for these calculations and may in fact even be overestimating the Orowan contributions in some of the samples.

Although the ZrC precipitates can significantly increase strength, their contribution fails to fully explain the increased hardness of these samples.

Grain boundary segregation has been shown to be another effective mechanism for increasing the strength of nanocrystalline metals. Vo et al. [75] studied doping of grain boundaries in nanocrystalline Cu with an average grain size of 8 nm through a combination of molecular dynamics and Monte Carlo simulations, in order to explore the effect of grain boundary segregation on grain boundary energy and yield strength. These authors observed a 60% increase in the yield stress after doping the material with only 1.2 at.% Nb, while also observing that the grain boundary energy is lowered with doping in a manner that is proportional to all observed strengthening. A similar observation was reported in molecular dynamics simulations of nanocrystalline Cu doped with Sb.[76] The nanoindentation hardness measurements of Ozerinc et al.[77] on nanocrystalline Cu-based alloys with an average grain size of 17 nm showed that doping grain boundaries with 10 at.% Nb and Fe can add an extra ~3 GPa and ~1 GPa to the hardness, respectively. These observations confirmed the previous molecular dynamic simulation findings that grain boundary segregation can significantly increase strength, showing a direct relationship between a decrease in excess grain boundary energy and an increase in strength. Such a relationship was also observed in nanoindentation studies by Rupert et al[78], who showed that accessing a lower grain boundary energy configuration with grain boundary relaxation can lead to higher hardness values in nanocrystalline Ni-W.

Consequently, the structure and excess energy of grain boundaries can be very important for determining the strength of nanocrystalline metals. Since the grain boundaries are the main source of dislocations in nanocrystalline metals, any process that facilitates or hinders grain boundary dislocation emission can decrease or increase the strength of the material. Although we dope our boundaries in traditional ways at lower temperatures, the formation of higher type grain boundary complexions should also affect interfacial energy and, therefore, strength. It is important to recall that complexions form at high temperatures because they are energetically favorable over a clean boundary, meaning they will reduce the excess energy of the system. In addition, there may be complex interactions of dislocations at the two newly created interfaces between the amorphous intergranular film and the grains. Brandl et al.[79] studied this situation with molecular dynamic simulations and found that these types of interfaces can attract dislocations during deformation, perhaps acting as sinks. In our study, we find that the "other" contribution to hardness, which is associated with segregation and boundary structure, becomes larger at higher annealing temperatures. Since we observe thicker amorphous intergranular films as temperature increases, it appears that complexions structure is also important for mechanical strength.

5. Conclusion

In this article, the evolution of grain size, grain boundary structure, and hardness of nanocrystalline Cu-3 at.% Zr during thermal annealing was studied. The following conclusions can be made:

- Significant thermal stability was observed, due to a combination of segregation of Zr atoms to grain boundary sites, complexion formation, and the formation of ZrC precipitates. Even annealing the alloy at the extreme condition of 98% of solidus temperature for a period of 1 week only caused coarsening an average grain size of 54 nm.

- A number of complexions states were observed, with higher temperatures promoting increased disorder and thicker grain boundaries, while lower temperatures gave rise to doped yet ordered interfaces. Amorphous intergranular films appeared above 850 °C, with higher temperatures promoting thicker films. While it is clear that the majority of these amorphous films were Type V and in equilibrium with the abutting grains, it is possible that a few were Type VI wetting films. Overall, amorphous complexions with thicknesses ranging from 0.5 to 5.7 nm were observed.
- The hardness of all of the samples remained higher than that of pure nanocrystalline Cu, with a maximum hardness value of 4.4 GPa measured. The combination of Hall-Petch and Orowan strengthening cannot fully explain the observed hardness, meaning that grain boundary segregation and complexion formation also lead to strengthening in this alloy. The influence of grain boundary state on strength becomes more significant at higher temperatures.

An important conclusion of this study is the formation of obvious complexions in a metallic alloy that retains its nanocrystalline grain size. Thick amorphous intergranular films were found at higher temperatures, giving a unique microstructure without the loss of strength associated with nanocrystalline metals. It was shown that by changing the annealing condition, different types of grain boundary complexions can be formed, providing a new set of tools for engineering the interfaces of nanocrystalline metals.

**Acknowledgement**

This study was supported by the U.S. Army Research Office under Grant W911NF-12-1-0511. Materials characterization was performed at the Laboratory for Electron and X-ray Instrumentation (LEXI) at UC Irvine, using instrumentation funded in part by the National Science Foundation Center for Chemistry at the Space-Time Limit (CHE-082913).

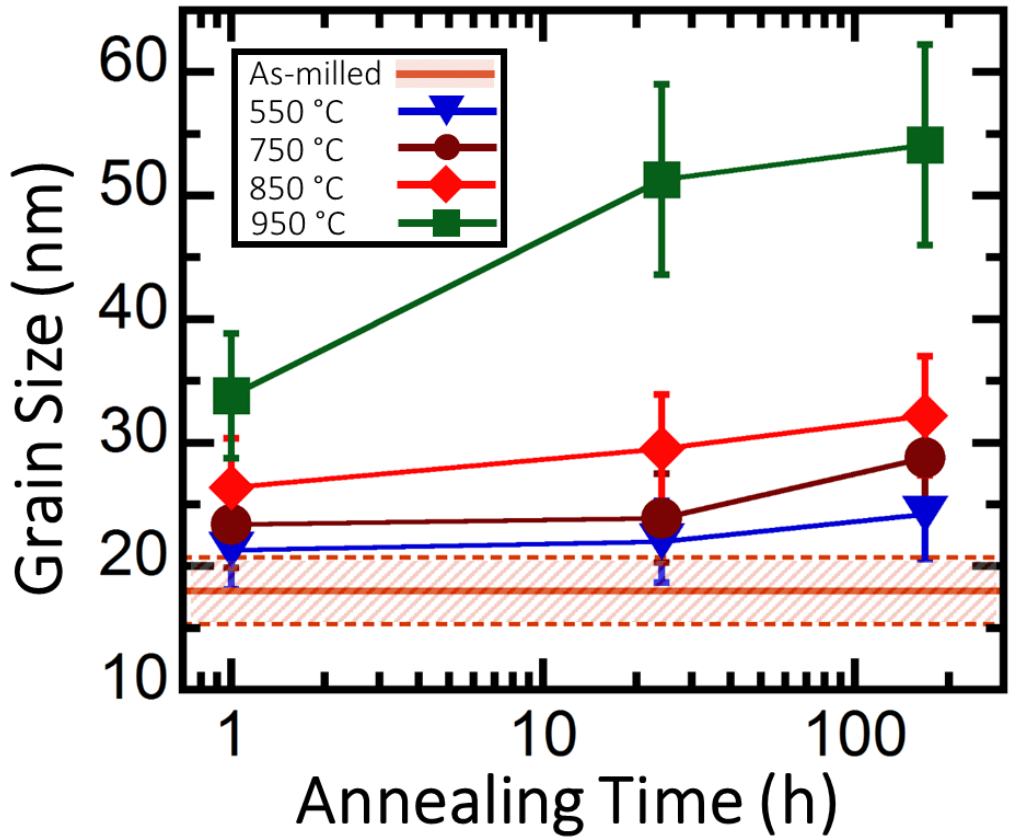

Fig. 1. XRD grain size for the as-milled Cu-Zr sample as well as samples annealed at different temperatures. A ±15% error bar was applied to all the measured grain sizes to account for measurement inaccuracies due to theoretical and instrumental limitations.

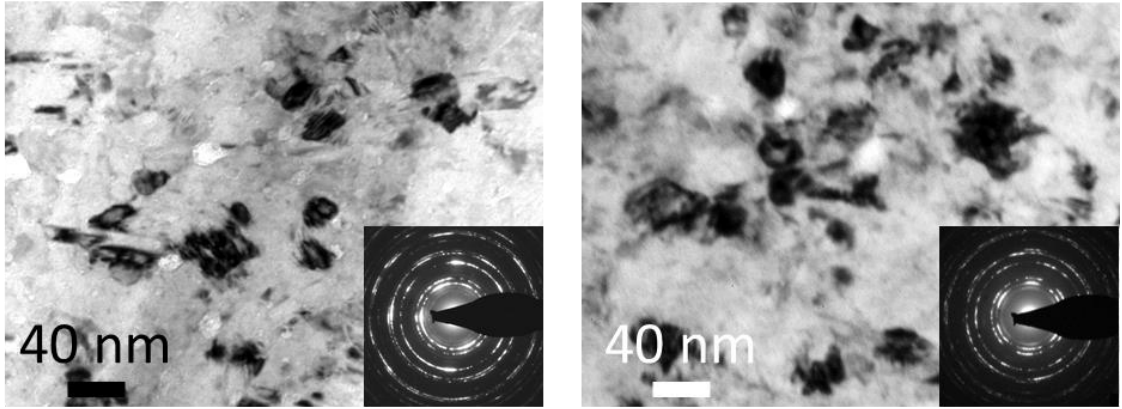
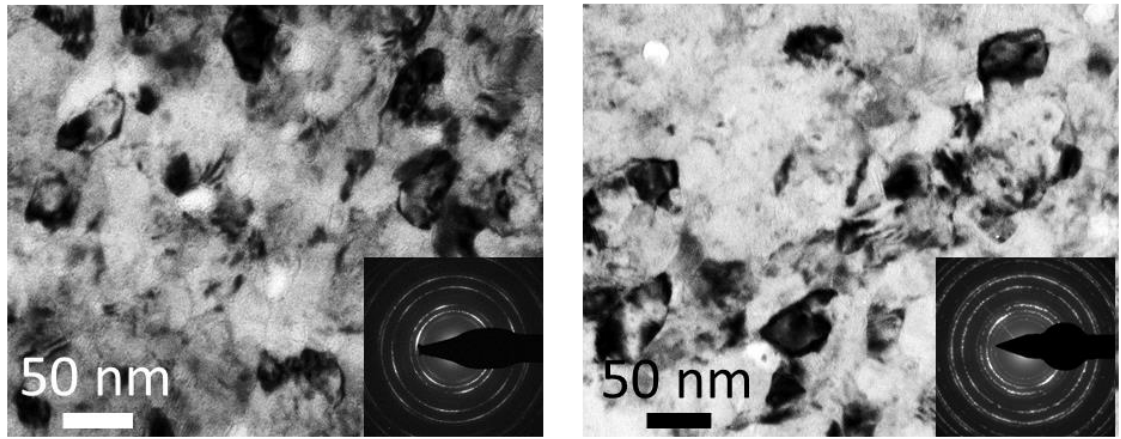

Fig. 2. Bright field TEM image of Cu-3 at.% Zr annealed at (a) 550 °C, (b) 750 °C, (c) 850 °C, and (d) 950 °C. The inset in each figure is the SAED pattern for that sample.

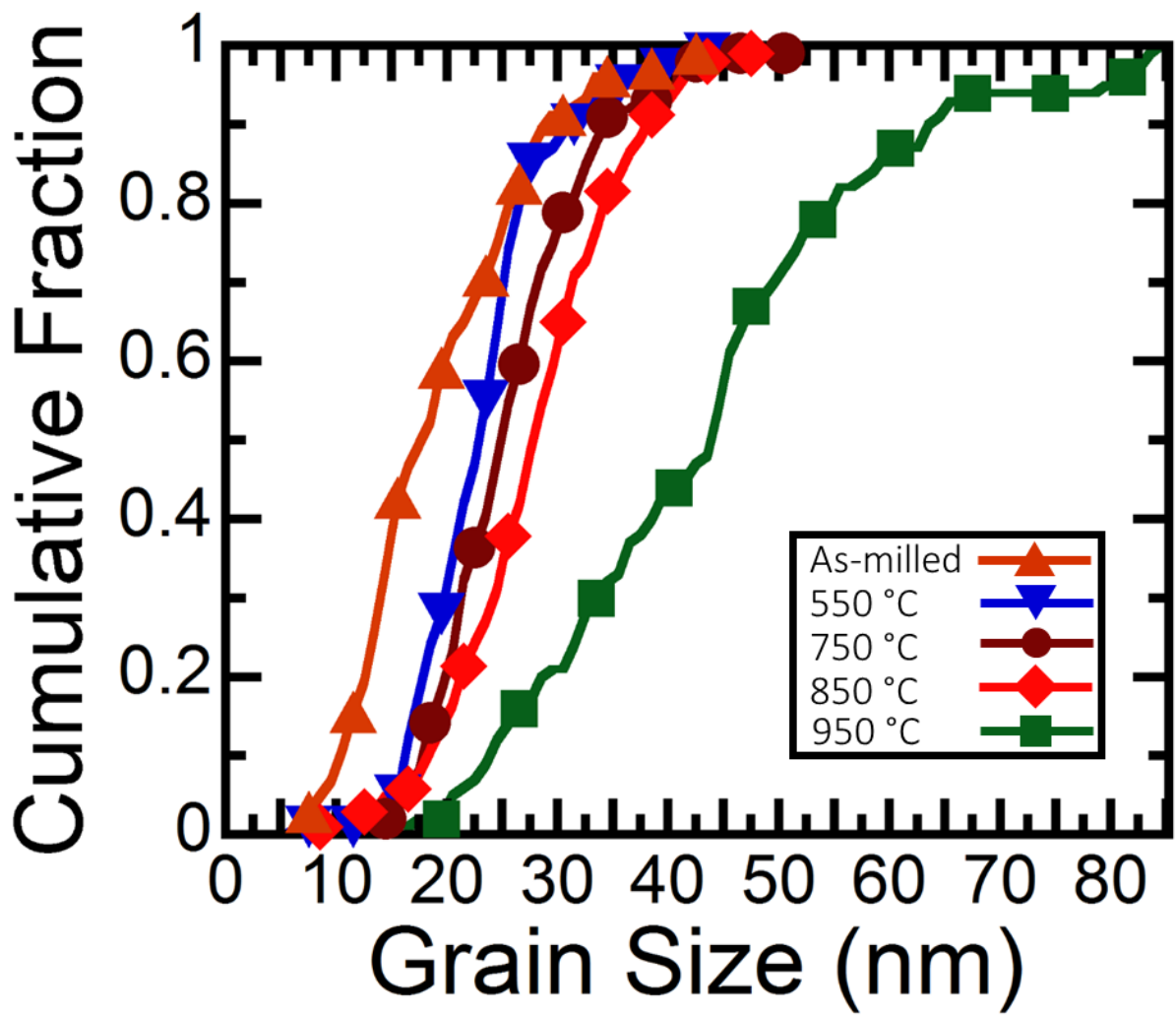

Fig. 3. Grain size distribution in the as-milled sample as well as samples annealed for 1 h at different temperatures. All the samples show a narrow grain size distribution, with the ratio of the standard deviation to the average being relatively constant.

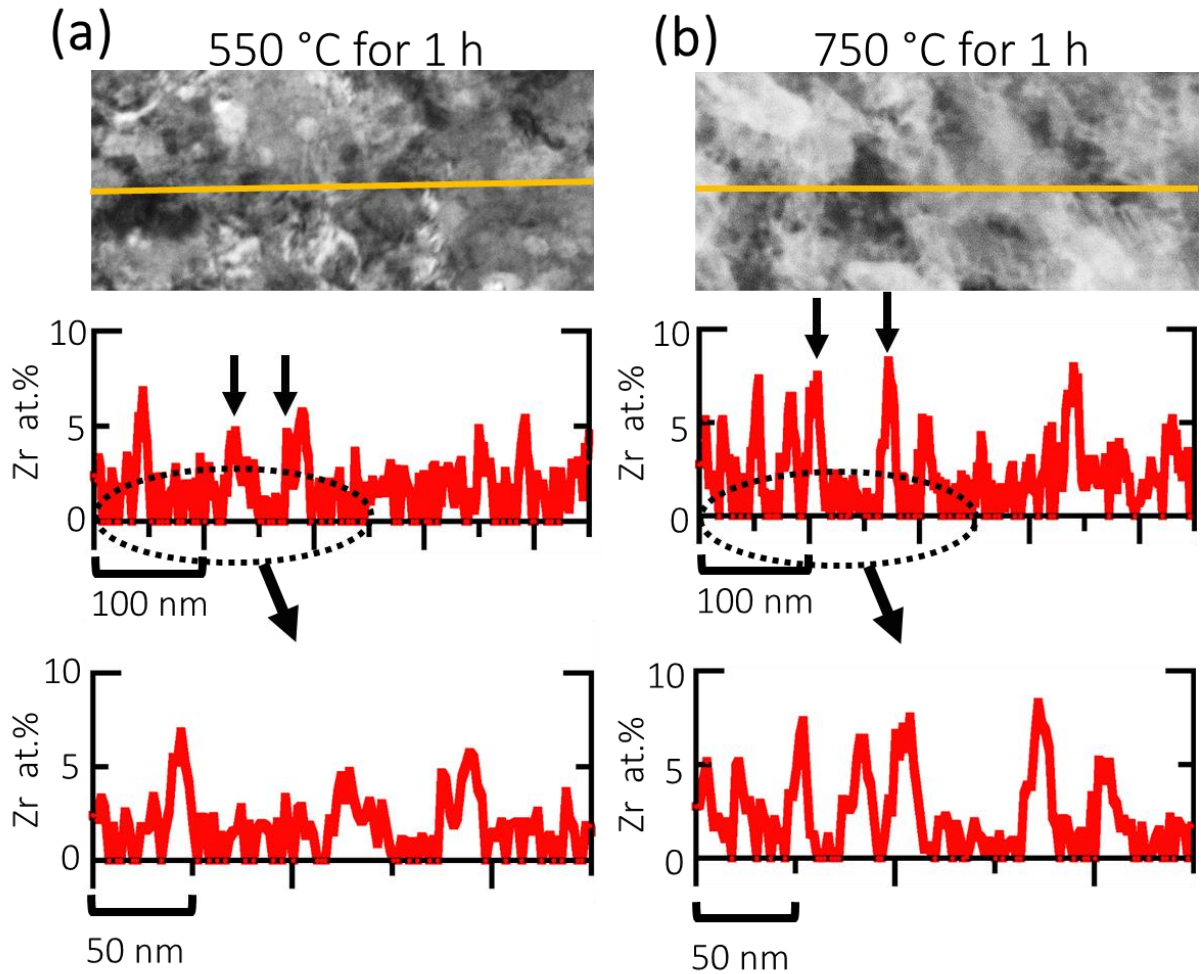

Fig. 4. Scanning TEM images and EDS line profile scans of the samples annealed for 1 h at (a) 550 °C and (b) 750 °C. A section of each profile from 0-250 nm was magnified and is shown directly below each figure. The locations of selected grain boundaries are marked with arrows.

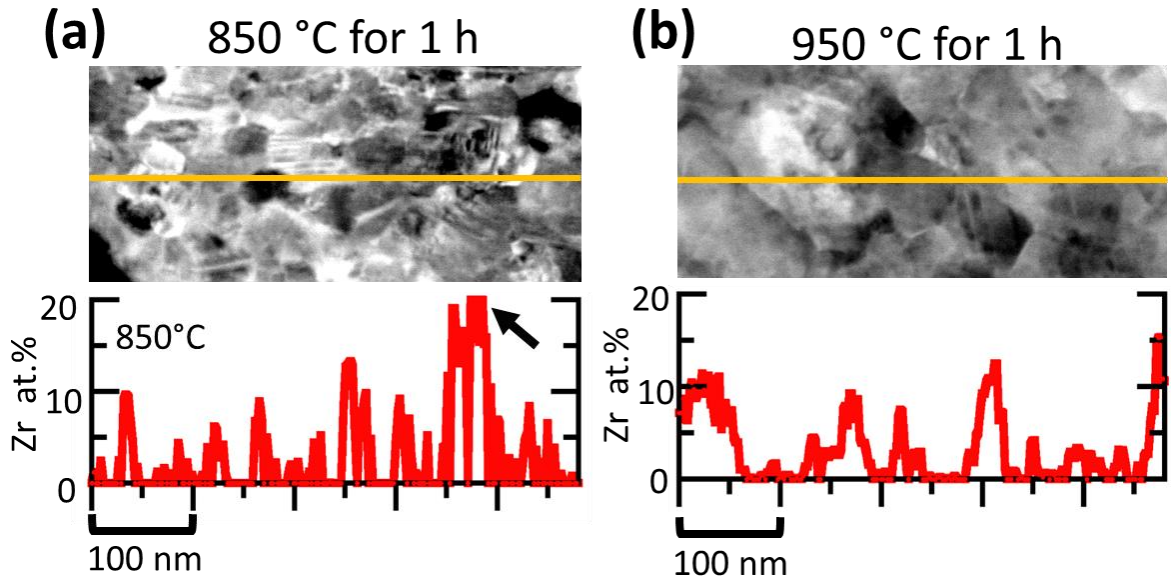

Fig. 5. Scanning TEM images and EDS line profile scans of the samples annealed for 1 h at (a) 850 °C and (b) 950 °C. An example of unusually high concentration of Zr in part (a) is marked with an arrow.

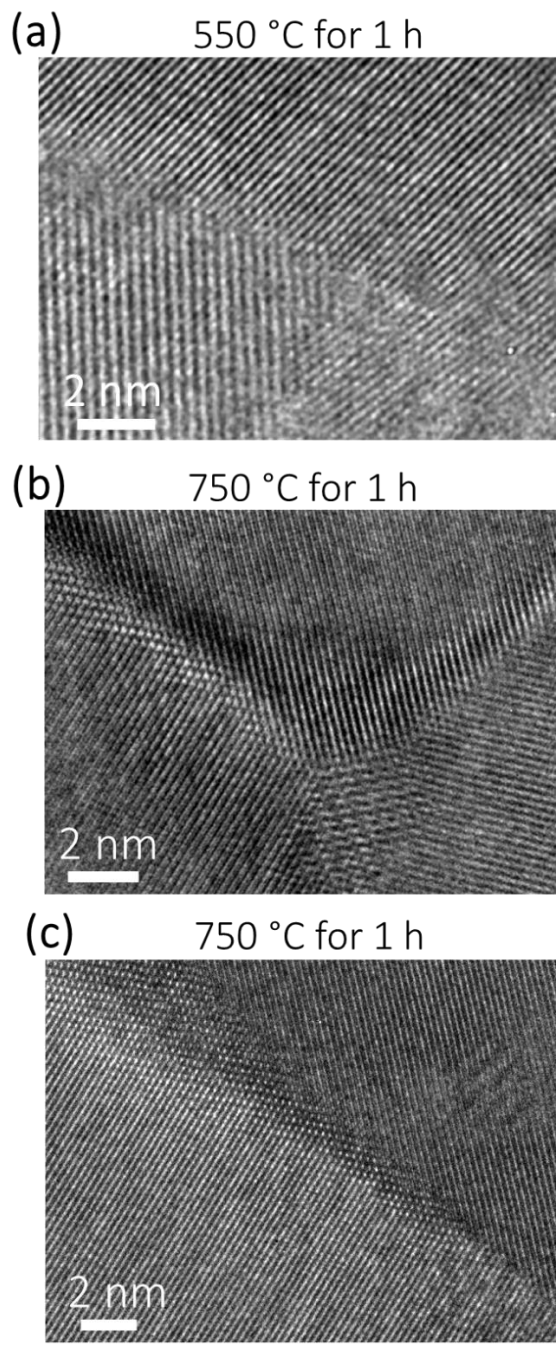

Fig. 6. High resolution TEM image of interfaces in Cu-3 at.% Zr. (a) A low energy configuration at the grain boundary in the sample annealed for 1 h at 550 °C. (b) A Fourier filtered high resolution TEM image of a fully connected triple junction. (c) A fully connected grain boundary in the sample annealed for 1 h at 750 °C.

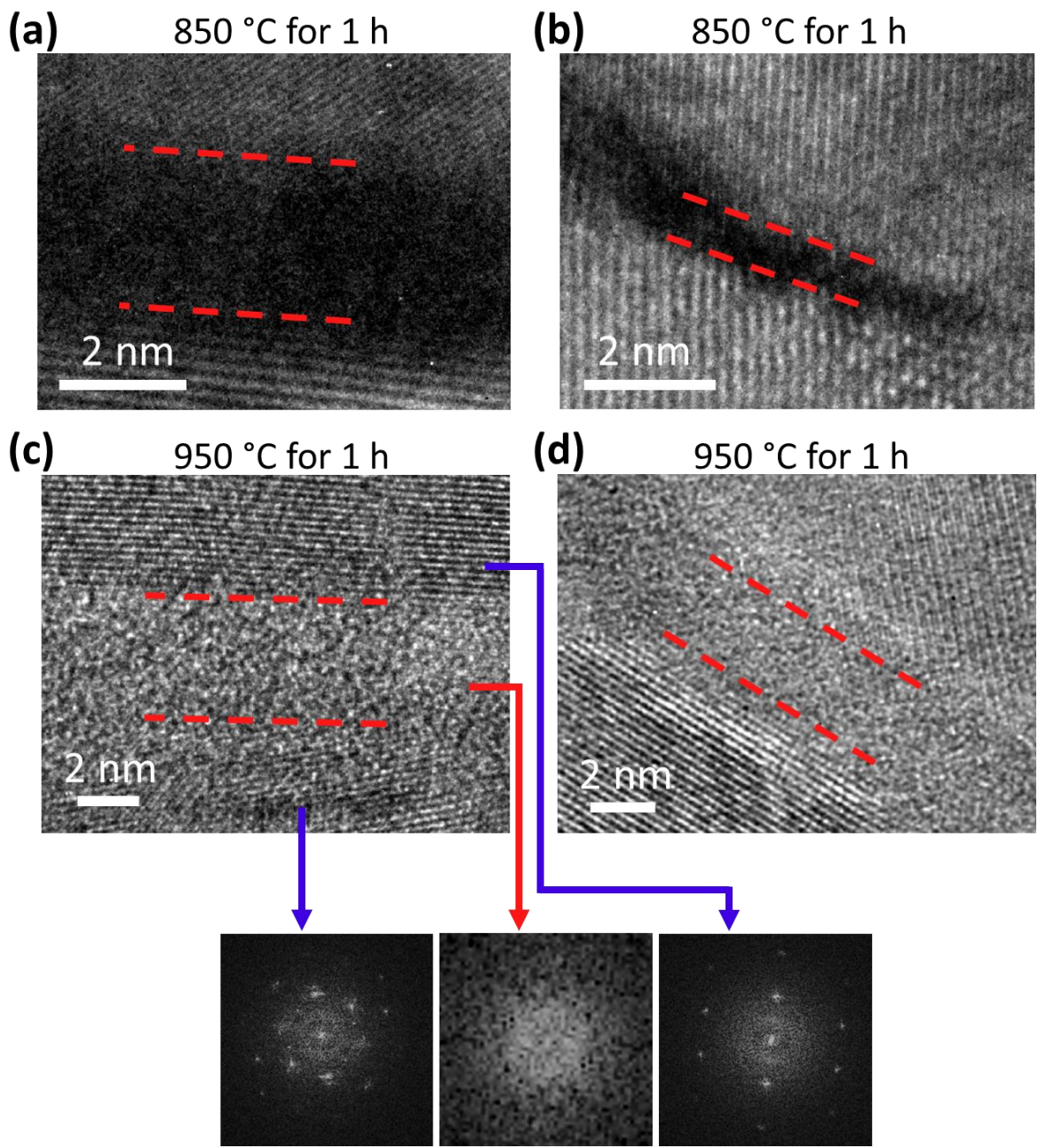

Fig. 7. High resolution TEM images of amorphous intergranular films in samples annealed at 850 °C and 950 °C with thicknesses of (a) 2.6 nm, (b) 0.8 nm, (c) 4.1 nm, and (d) 2.9 nm.

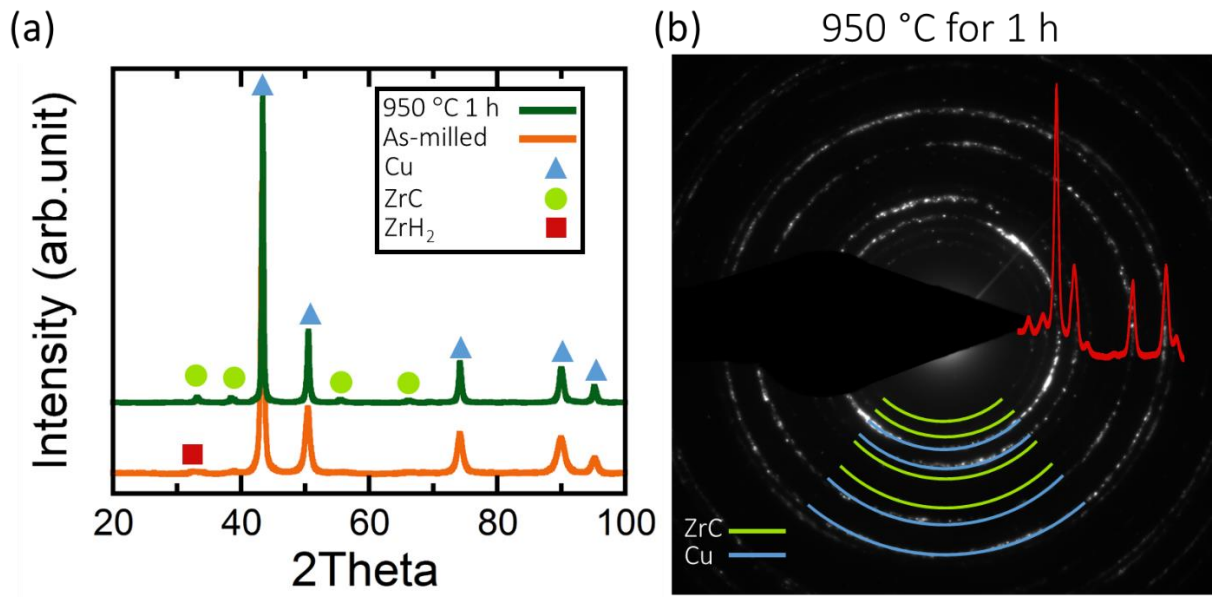

Fig. 8. (a) XRD pattern of the as-milled and 950 °C sample annealed for 1 h. The positions of Cu, ZrC, and $ZrH_2$ peaks are marked. (b) TEM SAED pattern of the sample annealed for 1 h at 950 °C with ZrC and Cu rings identified. An average intensity profile is also overlaid on the SAED pattern, using the PASAD script.[79]

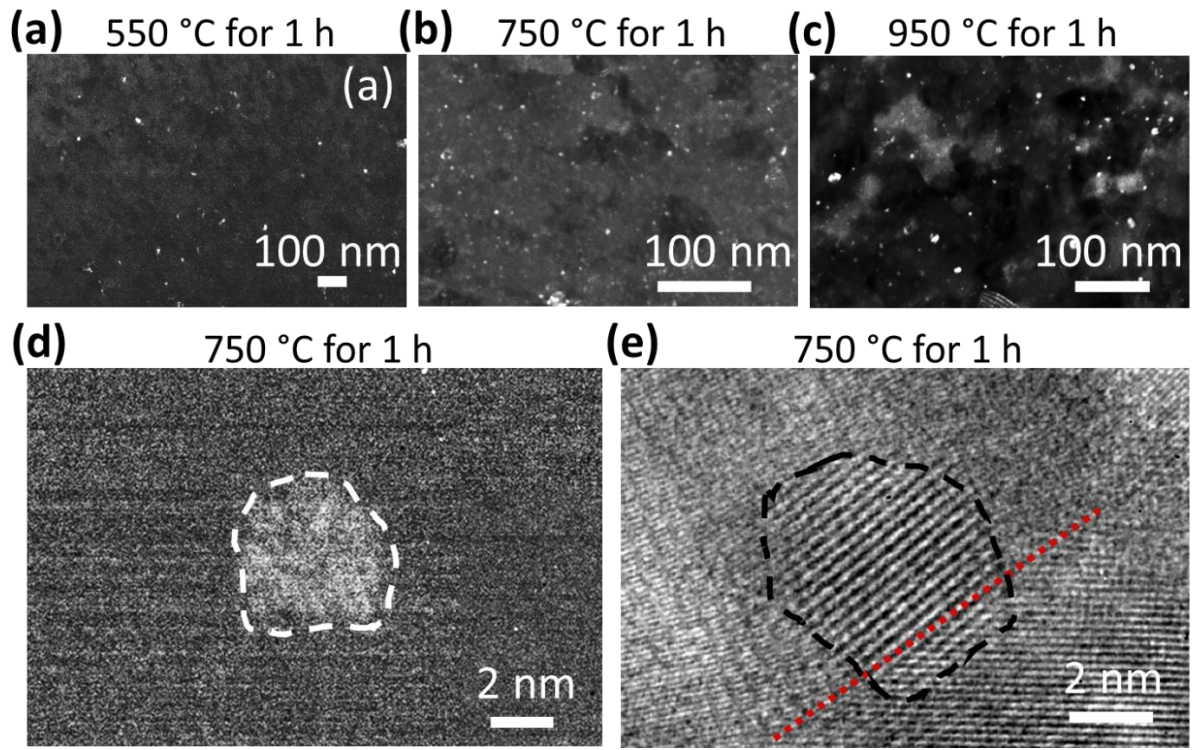

Fig. 9. Dark field TEM images of ZrC precipitates in the Cu-3 at.% Zr microstructure (a) annealed at 550 °C for 1 h, (b) annealed at 750 °C for 1 h, and (c) annealed at 950 °C for 1 h. (d) A higher resolution dark field TEM image of a ZrC particle. (e) The corresponding high resolution TEM image. The particle is sitting on the grain boundary, which is marked with a red dotted line.

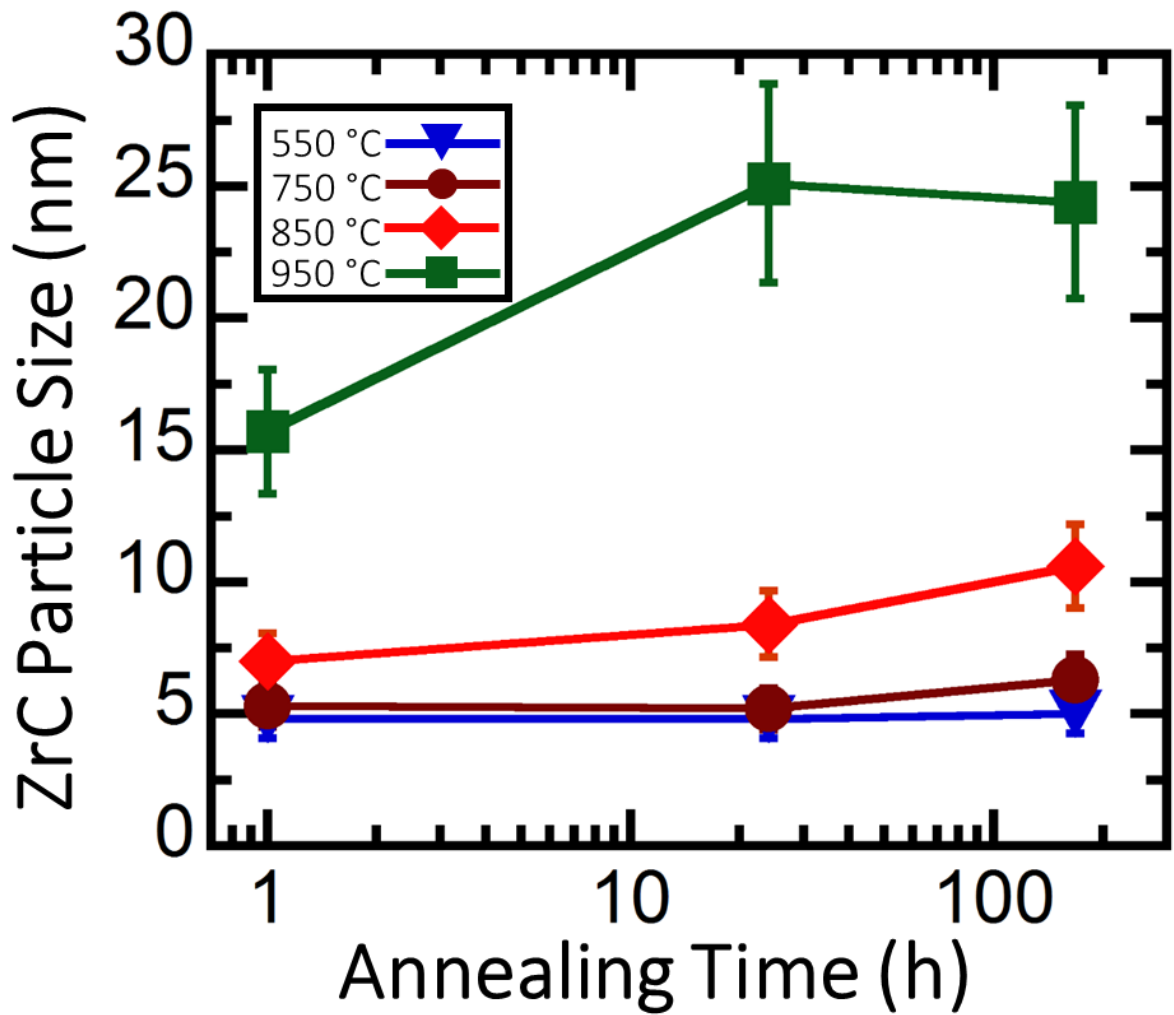

Fig. 10. ZrC particle size as a function of annealing time. The as-milled sample is not included in the graph since it does not contain ZrC particles.

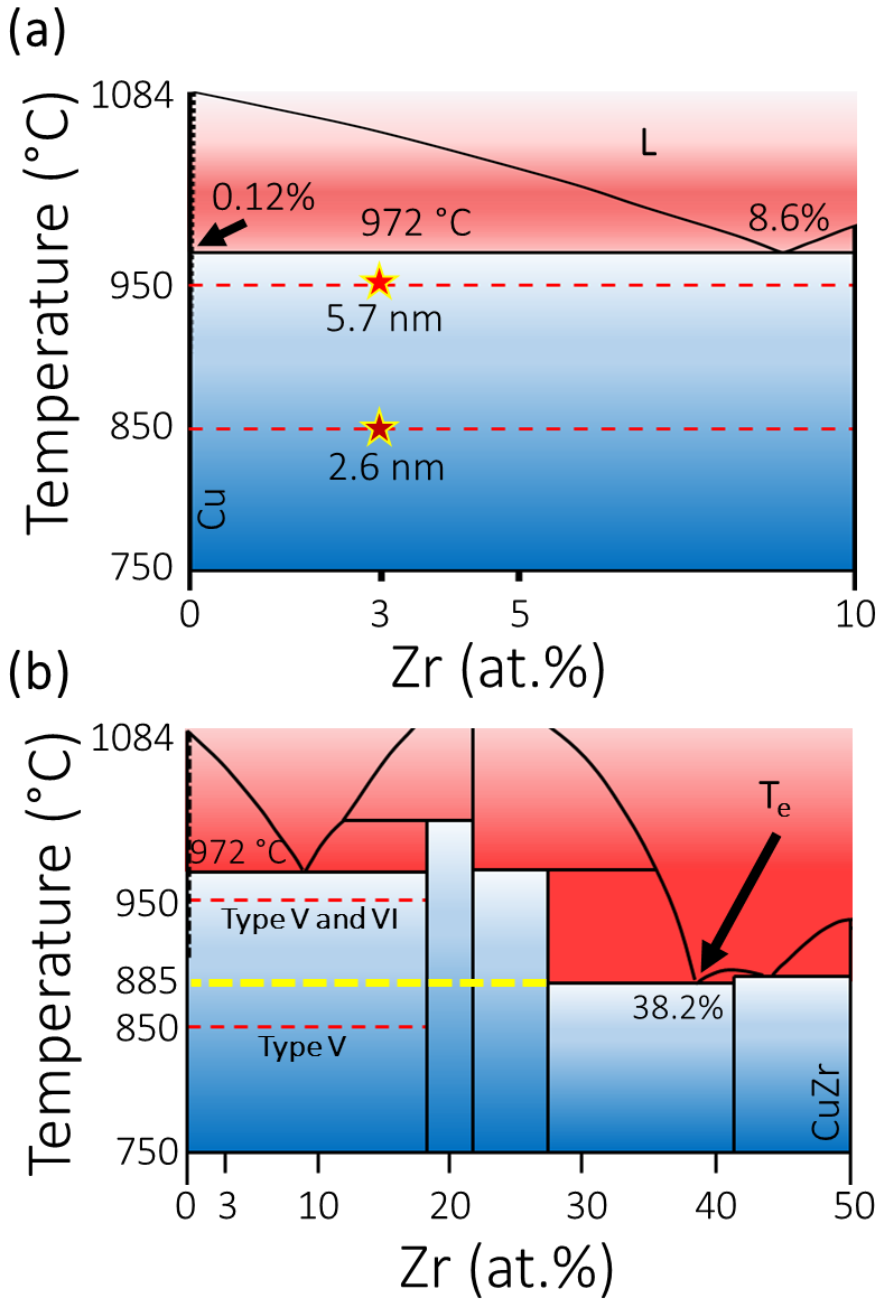

Fig. 11. (a) A section of Cu-Zr phase diagram, based on Ref. [64], showing the region of interest in this study. The dotted lines show the annealing temperatures where amorphous intergranular films were found, with the maximum film thickness labeled. (b) A larger section of the grain boundary phase diagram showing the lowest eutectic temperature and composition. The 850 °C sample is below all the solidus temperatures possible in this phase diagram, so any amorphous films must be complexion Type V. Blue shows region of all solid phase and red color shows the sections of all liquid or two phase solid-plus-liquid regions.

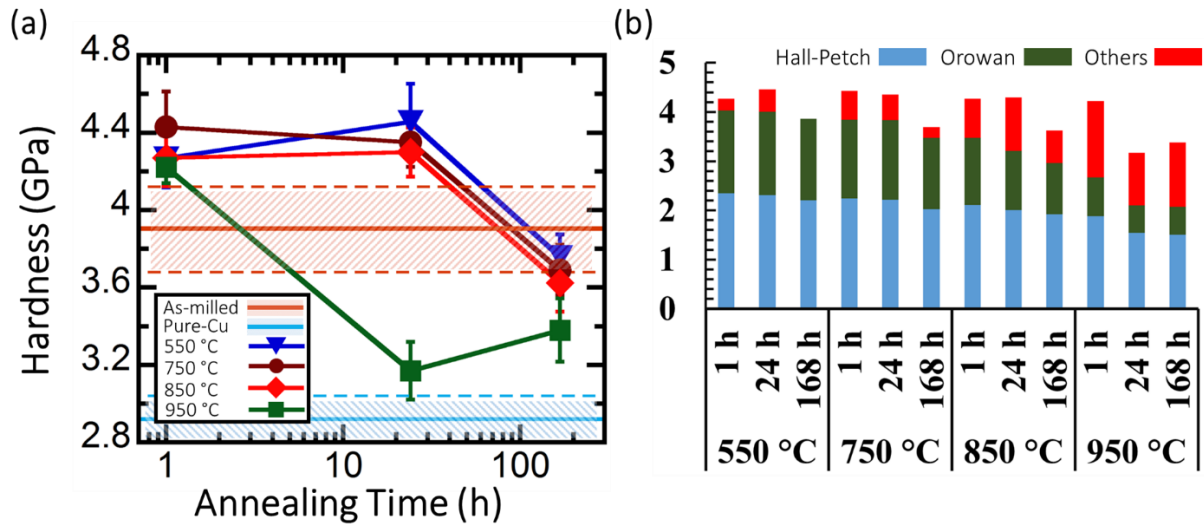

Fig. 12. (a) Nanoindentation hardness as a function of annealing time for nanocrystalline Cu and Cu-Zr alloys. (b) Stacked bar chart of Hall-Petch, Orowan, and "other" hardening mechanisms for nanocrystalline Cu-3 at.% Zr at different annealing times and temperatures.